\documentclass[runningheads]{llncs}
\usepackage{amsmath}
\usepackage{booktabs} 
\usepackage{caption} 
\usepackage{subcaption} 
\usepackage{graphicx}
\usepackage{pgfplots}
\usepackage[all]{nowidow}
\usepackage[utf8]{inputenc}
\usepackage{tikz}
\usetikzlibrary{er,positioning,bayesnet}
\usepackage{multicol}
\usepackage{algpseudocode,algorithm,algorithmicx}
\usepackage{hyperref}
\usepackage[inline]{enumitem} 
\usepackage{marvosym} 
\usepackage{xcolor}
\usepackage{graphicx}
\usepackage{fancyvrb}
\usepackage{float}
\usepackage{enumitem} 

\newcommand{\Rlogo}{\protect\includegraphics[height=1.8ex,keepaspectratio]{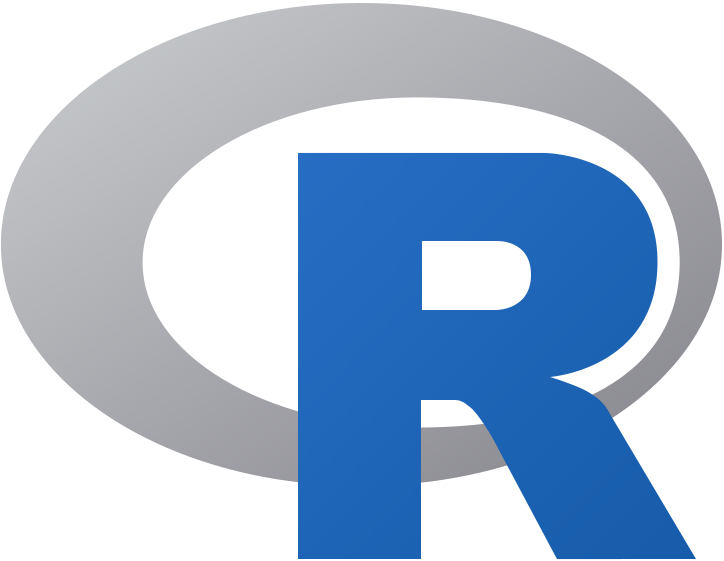}}
\definecolor{blue}{HTML}{1F77B4}
\definecolor{orange}{HTML}{FF7F0E}
\definecolor{green}{HTML}{2CA02C}
\definecolor{darkblue}{RGB}{37,37,195}
\definecolor{darkgreen}{RGB}{55,200,55}

\pgfplotsset{compat=1.14}

\begin{document}
\title{Simultaneous Confidence Corridors for neuroimaging data analysis: applications for Alzheimer's Disease diagnosis.}
\titlerunning{SCCs for Neuroimaging Data Analysis}
%

\author{Juan A. Arias \href{mailto:juanantonio.arias.lopez@usc.es}{\Letter} \textsuperscript{1}\and 
Carmen Cadarso-Suárez\textsuperscript{1} \and 
Pablo Aguiar-Fernández\textsuperscript{2,3}\vspace{0.5cm}}

\authorrunning{Arias et al. (2020) - Working draft}

%

\institute{
Biostatistics and Biomedical Data Science Unit. Department of Statistics, Mathematical Analysis, and Operational Research, Universidade de Santiago de Compostela, Spain \and Nuclear Medicine Department and Molecular Imaging Group, University Clinical Hospital (CHUS) and Health Research Institute of Santiago de Compostela (IDIS), Santiago de Compostela, Spain  \and
Molecular Imaging Group, Department of Psychiatry, Radiology and Public Health, Faculty of Medicine, Universidade de Santiago de Compostela, Spain
}


\maketitle 

\begin{abstract}

Alzheimer's disease (AD) is a chronic neurodegenerative condition responsible for most cases of dementia and considered as one of the greatest challenges for neuroscience in this century. Early AD signs are usually mistaken for normal age-related cognitive dysfunctions, thus patients usually start their treatment in advanced AD stages, when its benefits are severely limited. AD has no known cure, as such, hope lies on early diagnosis which usually depends on neuroimaging techniques such as Positron Emission Tomography (PET). PET data is then analyzed with Statistical Parametric Mapping (SPM) software, which uses mass univariate statistical analysis, inevitably incurring in errors derived from this multiple testing approach. Recently, Wang et al. (2019) \cite{Wang2019} formulated an alternative: applying functional data analysis (FDA), a relatively new branch of statistics, to calculate mean function and simultaneous confidence corridors (SCCs) for the difference between two groups' PET values. Here we test this approach with a practical application for AD diagnosis, estimating mean functions and SCCs for the difference between AD and control group's PET activity and locating regions where this difference falls outside estimated SCCs, indicating differences in brain activity attributable to AD-derived neural loss. Our results are consistent with previous literature on AD pathology and suggest that this FDA approach is more resilient to reductions in sample size and less dependent on \textit{ad hoc} selection of an $\alpha$ level than its counterpart, suggesting that this novel technique is a promising venue for research in the field of medical imaging.

\keywords{Neurodegenerative disease \and Alzheimer's disease \and Medical imaging \and Neuroimage \and Positron emission tomography \and Functional data analysis  \and Simultaneous confidence corridors \and Bivariate splines}
\end{abstract}


\section{Introduction} \label{sec:Introduction}

Motivated by Wang et al.'s \cite{Wang2019} work, where they propose the use of functional data analysis (FDA) to the field of medical imaging, in this article we carry out a practical application by calculating simultaneous confidence corridors (SCCs) of positron emission tomography (PET) data obtained from Alzheimer's disease (AD) and control (CN) patients. We apply this FDA procedure both for one-sample and two-sample setups, then visually compare results with the ones obtained using Statistical Parametric Mapping (SPM), and assess the potential utility of this novel technique for AD diagnosis. While a detailed review of AD pathology and the calculation of SCCs is out of scope for the current paper, we aim to give a sufficiently consistent background for all researchers interested in this subject and to provide computational tools to ensure the replicability of this study. 


\subsection{Alzheimer's Disease} \label{subsec:AD}

AD is the most common neurodegenerative disease (NDD) and sole responsible for the majority of cases of dementia \cite{facts}, as well as a major death cause \cite{Helmer2001}. NDD is an umbrella term which describes a wide number of progressive, incurable, and highly debilitating medical conditions caused by neural degeneration \cite{Przedborski2003}. NDD are often age-related, as such, vulnerable population to these diseases is expected to increase substantially in the following decades, with estimates predicting that people over 60 years old will double between the year 2000 and 2050 \cite{world2015world}. By that year, patients suffering from dementia will exceed 115 million worldwide \cite{prince2015world}. This growing problematic, known as the burden of neurodegenerative diseases \cite{Fereshtehnejad2019}, is considered one of the greatest challenges for neuroscience in this century, as it threatens not only patients, but also their families, caregivers, and public health systems' capacities.

Risk factors for AD include age, smoking, heart disease, hypercholesterolemia, hypertension, alcohol consumption, low physical activity, and diabetes \cite{patterson2008,Rosendorff2007,Mayeux2012}. On the other hand, factors hypothesized to prevent AD include physical exercise \cite{Podewils2005}, a healthy diet \cite{Scarmeas2006}, and high cognitive reserve \cite{xu2019association}, among others \cite{patterson2008,Mayeux2012}. There are two types of AD depending on onset age: early-onset AD (i.e. familial AD), accounting for 1\% of the cases \cite{Goedert2006}; and late-onset AD (LOAD), accounting for the remaining 99\% of cases. Familial AD causes are genetic and easy to determine (e.g. the Osaka mutation \cite{Tomiyama2020}), but the situation is much less clear for LOAD, with low heritability rates \cite{Wilson2011} and a variety of 19 loci identified which are associated with AD but still leaving a large portion of genetic risk without explanation \cite{Lambert2013}.

\begin{figure}[h!]
\centering 
\vspace{0.3cm}
\subfloat[Coronal view]{%
  \includegraphics[clip,width=0.3\columnwidth]{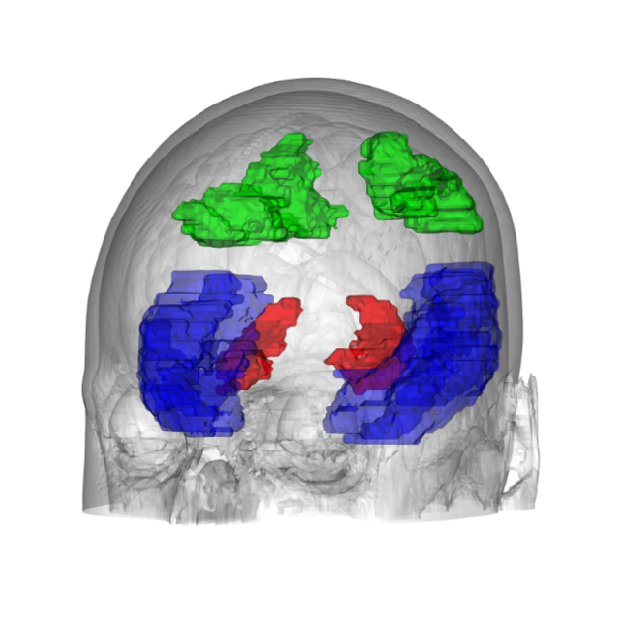}
}
\subfloat[Sagittal left view]{%
  \includegraphics[clip,width=0.3\columnwidth]{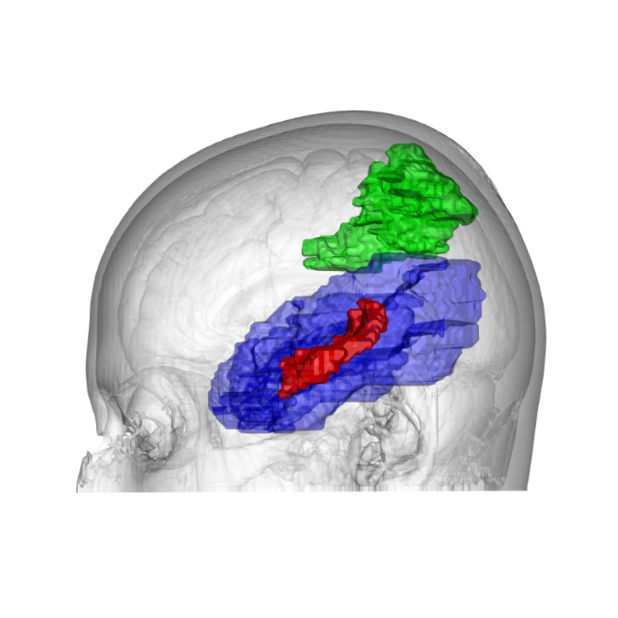}
}
\subfloat[Axial superior view]{%
  \includegraphics[clip,width=0.3\columnwidth]{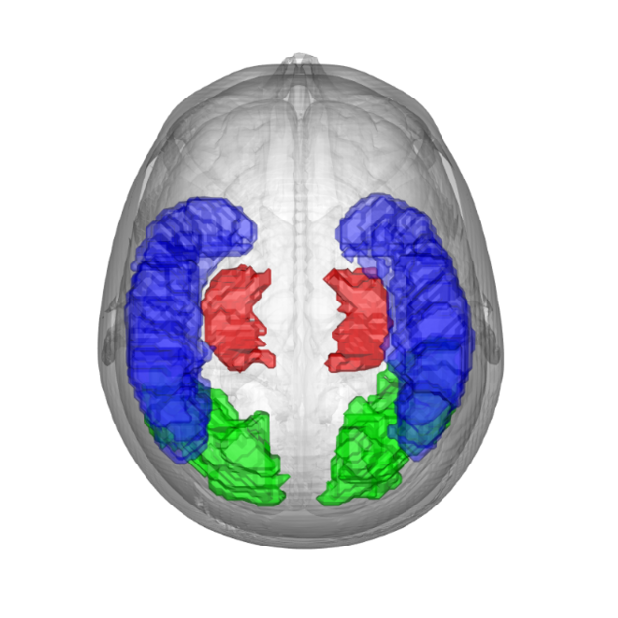}
}
\vspace{0.3cm}
\caption{Regions of interest for research on early AD diagnosis in the human brain: hippocampus (\textcolor{red}{red}), temporal cortices (\textcolor{darkblue}{blue}), and parietal cortices (\textcolor{darkgreen}{green}).}
\vspace{0.3cm}
\label{fig:location}
\end{figure}

AD main effect is progressive neurological damage resulting from neuronal loss which occurs first in the hippocampus, temporal cortices, and parietal cortices \cite{McKhann1984,Wenk2003} (see Figure \ref{fig:location}). As the disease evolves, neural damage progresses and affects the parahippocampal gyrus and amygdala \cite{Mahanand2012}. In advanced stages, AD affects brain regions involved in primary bodily functions, complicating vital movements and becoming fatal to the patient \cite{facts}. Caused by the presence of neuritic plaques \cite{Iwatsubo1994} and neurofibrillary tangles \cite{Clark1998} in the affected neurons, this progressive neurological damage translates into neuropsychological and behavioural manifestations including worsened capacity to recall information, problems with planning and decision-making, generalized confusion, delusions, hallucinations, repetition of conversations, anxiety, among others \cite{Jalbert2008,facts}. In addition, AD is typically accompanied by comorbidities such as major depressive disorders \cite{Lyketsos2002} which complicate the already poor situation of the patient, aggravating perceived health in a self-sustained downward spiral as proposed by recent life-course models \cite{Kemp2017} and contributing to severe reductions in life expectancy and other diffuse effects on health and well-being \cite{Arias2020}. 

Despite our knowledge of AD neuropathology, no theoretical framework can explain the formation of these plaques and tangles, nor the specific regional distribution displayed. A number of coexisting and non-exclusive hypothesis have been proposed, including the amyloid cascade hypothesis \cite{Hardy2006}, the cholinergic hypothesis \cite{Francis1999}, or the Tau hypothesis \cite{Maccioni2010}, among others \cite{Moulton2012,Sochocka2017}.

AD diagnosis is truly challenging as the pathological process may start between 20 to 30 years before the first detectable symptoms appear \cite{Sperling2014,Goedert2006}. In addition, early AD stages are very similar to the common forms of senile dementia \cite{Terry1980} and mild cognitive impairment (MCI) \cite{Ward2013}, increasing the complexity of a precise early diagnosis. Several biomarkers have been proposed \cite{Jalbert2008,Tan2014}, but they all suffer from limitations and a precise diagnosis usually happens in late AD stages by means of test batteries such as the Mini Medical State Examination \cite{Jalbert2008} based on standard diagnostic criteria (i.e. DSM-IV-TR) \cite{AmericanPsychiatricAssociation2013}. 
Another option for AD diagnosis is the use of neuroimaging tools such as magnetic resonance imaging (MRI) \cite{Jack2002} and single-photon emission computed tomography \cite{McNeill2007}. PET, a neuroimaging technique which measures regional metabolic rates of radioactively-marked glucose, has been applied to AD diagnosis in early developmental stages \cite{Silverman2002,Mosconi2013,Marcus2014,Nordberg2010} and for differential AD diagnosis \cite{zamrini2004imaging}. A variant of PET, 18F-FDG PET, is the technique used in our study; further details are discussed in Section \ref{subsec:PET}.

With regards to treatment, current pharmacotherapeutic interventions for AD are carried out with drugs such as cholinesterase inhibitors \cite{birks2006cholinesterase} and memantine \cite{robinson2006memantine} in order to slow down the progress of the disease. However, results have been modest, difficult to translate into generalized medical benefit \cite{Jalbert2008}, and limited to only five medical treatments for AD approved by 2016 \cite{Briggs2016}. In addition, these treatments are designed as symptomatic rather than disease-modifying and it is not probable that a single cure for AD will appear \cite{Briggs2016,Mangialasche2010}. Non-pharmacoterapeutic treatments are also applied although the evidence on their efficiency is incomplete and not lacking contradictions \cite{Dietch1989}.

Given the growing problematic of AD in modern societies and the limited success of therapeutic interventions, there is a growing interest for early AD diagnosis considering that available treatments are more beneficial in initial stages of the pathological process \cite{Cummings2007,Reiman2016} and that AD-derived neurodegeneration starts between 20 to 30 years prior to the appearance of the first clinical symptoms \cite{Goedert2006,Sperling2014}. Thus, early diagnosis of AD will be critical to reduce the impact of this disease in our societies, especially bearing in mind that - as an age-related condition - small delays on disease onset are expected to markedly reduce AD prevalence \cite{Briggs2016,Brookmeyer2007}. Besides, early treatment would slow down or even stabilize the disease, reducing the need for external care, improving patients' life quality, and increasing a life expectancy which is currently estimated to range between three to ten years after diagnosis \cite{Zanetti2009}. After this review of the current state of the art on AD, we consider that early diagnosis is the global imperative for researchers in order to reduce the impact of this upcoming pandemic.


\section{Objectives} \label{sec:objectives}

Having reviewed basic concepts on AD and considering the imperative need for early diagnosis of this severe condition, the proposed aim of this article is to evaluate potential improvement for AD diagnosis by means of FDA techniques, applying SCCs to the difference between mean functions of PET data from AD and CN groups and examining regions exhibiting differential patterns of activity. In order to achieve our aim we propose the following objectives:

\begin{enumerate}[label=(\alph*),leftmargin=2\parindent, rightmargin=1\parindent]
    \vspace{0.25cm}
    \item To calculate SCCs for one-sample and two-sample cases using bivariate splines over triangulations of 18FDG-PET data.
    \vspace{0.25cm}
    \item To visualize areas where the difference between estimated mean functions falls outside estimated SCCs.
    \vspace{0.25cm}
    \item To evaluate whether obtained results are in line with existent scientific literature on regional distribution of AD-derived neuronal loss.
    \vspace{0.25cm}
    \item To visually compare these results with the ones obtained using SPM software under the same conditions in order to visually assess performance of proposed FDA methodology.
\end{enumerate}{}


\section{Materials and Methods} \label{sec:data}

\subsection{Brain Imaging Data} \label{subsec:PET}

There are slightly different approaches for PET imaging depending on the radioisotope used; in 18F-FDG PET, Fluorodeoxyglucose (18F-FDG), a radioisotope analog of glucose, is used as tracer to monitor brain metabolic rates. Positron emission rates by molecules of 18F-FDG trapped in brain tissues are used as an indirect measure of glucose consumption, which is then reconstructed \cite{Leahy2000} producing 3D images for the position of this tracer in the brain.

18F-FDG PET has been recently included in AD diagnostic routines and is the methodology chosen for our study. We use  Statistical Parametric Mapping (SPM) \cite{Penny2011} software to carry out anatomical standardization which allows for inter-subject comparisons after a series of pre-processing stages. This step is key in any neuroimaging study as the existence of a precise point-to-point correspondence is critical when comparing scans from brains which present characteristic shape and size. For this reason, we rely on SPM's anatomical standardization routines as they have been thoroughly developed and tested \cite{Ishii2001}. 

\begin{figure}[h!] 
  \centering 
  \vspace{0.5cm}
  \begin{subfigure}[b]{0.32\linewidth}
    \includegraphics[width=\linewidth]{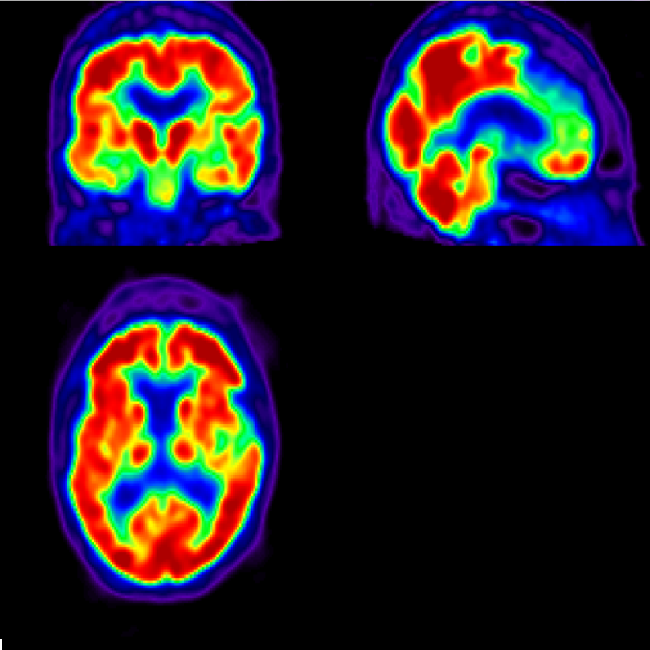}
    \caption{Raw image}
  \end{subfigure}
  \begin{subfigure}[b]{0.32\linewidth}
    \includegraphics[width=\linewidth]{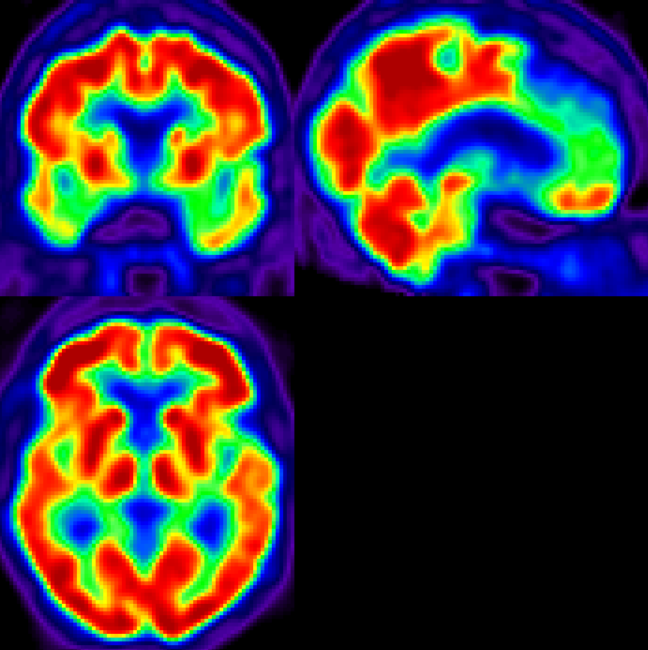}
    \caption{Normalized image}
  \end{subfigure}
  \begin{subfigure}[b]{0.32\linewidth}
    \includegraphics[width=\linewidth]{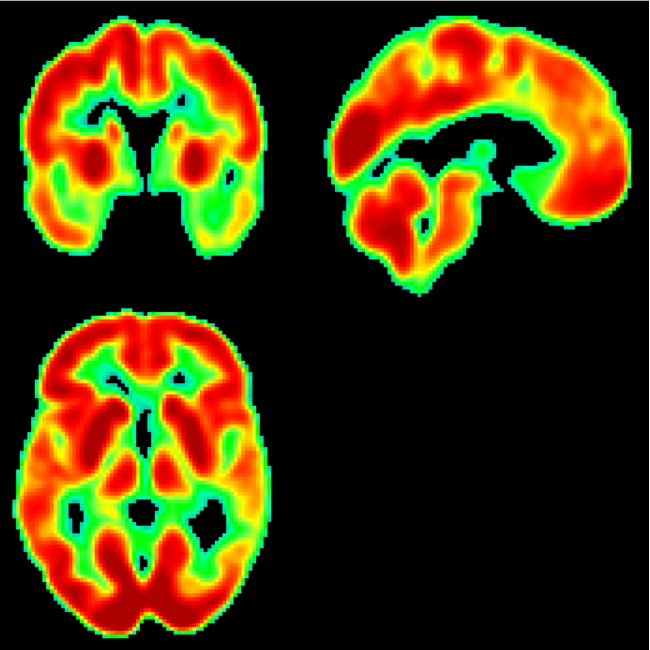}
    \caption{Masked image}
  \end{subfigure}
  \vspace{0.3cm}
  \caption{Stages of 18F-FDG PET processing. (a) Raw image obtained from ADNI, (b) image after anatomical standardization, (c) processed image after masking.}
  \vspace{0.5cm}
  \label{fig:maskedPET}
\end{figure}

In order to obtain 18F-FDG PET data we drawn upon the Alzheimer's Disease Neuroimaging Initiative (ADNI) \cite{Mueller2005}, a platform that collects data from different research institutions focusing on AD diagnosis. We selected 18F-FDG PET and MRI data, together with demographic information (age and sex) for CN group (75 patients; 44 male; age: $75.56 \pm 4.96$ years) and AD group (51 patients; 30 male; age: $74.03 \pm 7.25$ years) summing 126 participants. Images were then realigned, unwrapped, co-registered with MRI data, spatially normalised, mean proportionally scaled, and masked following standard procedures deployed by SPM with the aim of guaranteeing voxel-to-voxel comparability between images (see Figure \ref{fig:maskedPET}).


The resulting 18F-FDG PET image files were then converted into a regular data frame for its use in \Rlogo. The result is a data frame where every 18F-FDG PET value is pooled together with its correspondent coordinates in the 3D brain map and its corresponding demographic information and study group. Details for these bulk transformations using Matlab and \Rlogo \thinspace \thinspace can be found and replicated using the scripts stored in our \href{https://github.com/iguanamarina/SCCsneuroimage}{\textcolor{blue}{open code repository}}.  


\subsection{Simultaneous Confidence Corridors} \label{subsec:SCCs}

In a broad sense, FDA can be defined as the field of statistics which deals with the theoretical foundations and necessary tools for the analysis of data expressed in the form of functions, which can be extended to data in the form of images and shapes. FDA has experimented a great rise in popularity in the last decades and comprehensive publications on this field - including monographs \cite{ramsay2005principal,ferraty2006nonparametric} and review articles \cite{wang2016functional,ullah2013applications} - are now published explaining FDA's theoretical basis and applicability. Recently, Wang and colleagues \cite{Wang2019} proposed a novel approach to obtain mean values of imaging data and their corresponding SCCs (also known as 'simultaneous confidence bands') using FDA and thus avoiding typical issues of brain activity estimation such as the leakage in complex data structures and the multiple comparison problematic. In their article, the authors prove this method to be statistically consistent and asymptotically normal under standard regularity conditions. 

\begin{figure}[h!] 
  \centering 
  \vspace{0.25cm}
  \begin{subfigure}[b]{0.32\linewidth}
    \includegraphics[width=\linewidth,height=5cm]{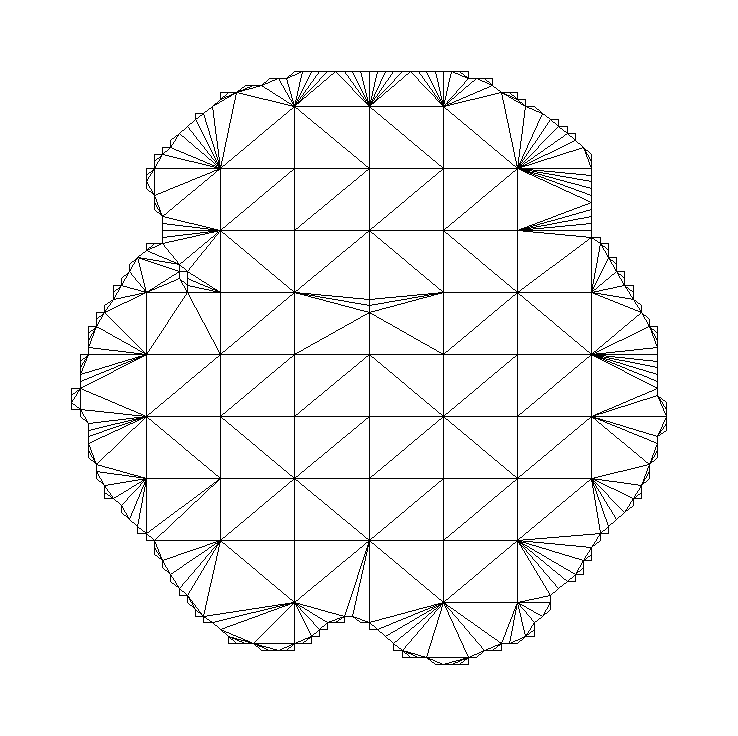}
    \caption{N=8}
  \end{subfigure}
  \begin{subfigure}[b]{0.32\linewidth}
    \includegraphics[width=\linewidth,height=5cm]{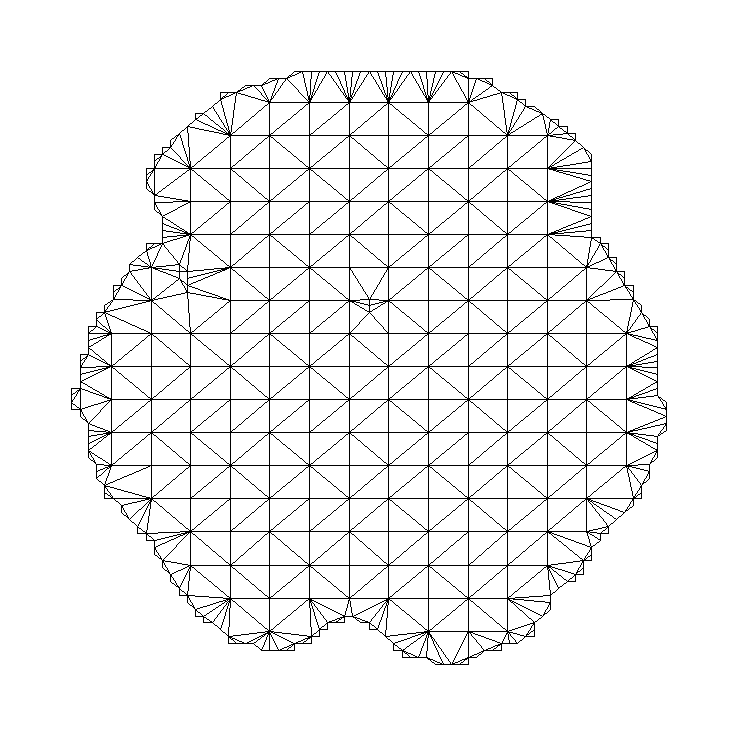}
    \caption{N=15}
  \end{subfigure}
  \begin{subfigure}[b]{0.32\linewidth}
    \includegraphics[width=\linewidth,height=5cm]{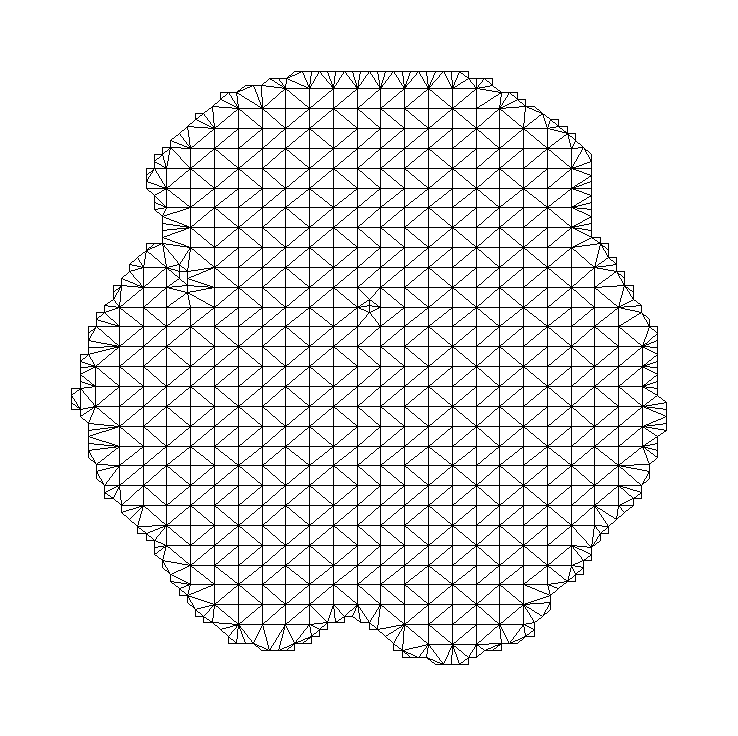}
    \caption{N=25}
  \end{subfigure}
  \vspace{0.4cm}
  \caption{Delaunay triangulations for SCCs construction at $Z=30$. Increasing N values represent increases in triangulation's degree of fineness. Wang et al. \cite{SCCpack} consider a fineness degree of N=8 as sufficient.} 
  \label{fig:triangulations}
  \vspace{0.5cm}
\end{figure}

First, images used in biomedical studies are often irregularly shaped, which implies a need for reliable smoothing methods. However, such methods usually suffer from a problem of \textit{leakage} with complex data structures such as the ones here considered, showing difficulties in the estimation for boundary regions. With our FDA approach, this problem is addressed by applying bivariate splines over Delaunay triangulations \cite{Chew1989}, thus preserving the features of these complex brain images \cite{Lai2013} (see Figure \ref{fig:triangulations}).


Following, when analyzing medical imaging data, problems arise not only for the estimation of the mean value for a given point, but also for the estimation of associated uncertainty, a problem which gets even more complex when considering the spatial correlation between points. Usually, the predominant analytic techniques deployed in software such as SPM follow the \textit{mass univariate approach} which considers pixels as independent units and then compares them with classical methods such as a \textit{T}-test. The multiple comparisons problem is then addressed with approaches such as the Bonferroni correction or the random field theory \cite{Worsley2004}. However, these are \textit{ad hoc} corrections and depend on the choice of an appropriate threshold, which is based on the designated professional's expertise. Wang et al.'s proposal considers data as functional, that is, as the result of a continuously defined function observed on a pre-defined regular grid. As such, attention moves from using pixels as basic analytic units to considering images as a whole and then calculating corresponding SCCs. The utility of this approach for the estimation of confidence bands for functional data has been previously addressed in the scientific literature \cite{Degras2011,Choi2018}. 

Here we propose the use of this FDA methodology to a practical case using PET data from CN and AD patients of different sexes and age ranges. Our goal is to show how this method can be applied to neuroimaging data and accurately point out regional differences in neural activity between two groups. We used \Rlogo \thinspace \thinspace language together with packages \href{https://cran.r-project.org/web/packages/triangulation/index.html}{Triangulation} \cite{Triangulation}, to get custom Delaunay triangulations for the boundaries of our data, and Wang et al.'s \href{https://rdrr.io/github/funstatpackages/ImageSCCs/}{ImageSCC} \cite{SCCpack} package, to calculate estimated mean functions and corresponding SCCs. This analysis is then replicated filtering by sex and age ranges. Results are shown in the following section, together with a succinct explanation of what can be observed in the attached figures and a visual comparison using a classic SPM approach under the same conditions.


\section{Results} \label{sec:results}
 
In this section we describe results obtained using the aforementioned FDA technique to PET data from CN and AD groups. First, in the one-sample case we calculate estimated mean functions and SCCs for each of the two groups individually. Then, in the two-sample case, we carry out comparisons between AD and CN entire groups and replicate them filtering by sex and age. These comparisons determine which brain areas have PET activity levels falling inside our estimated SCCs for the difference between groups, suggesting normal levels of brain activity, and which areas fall outside estimated SCCs for the difference between groups, suggesting hypo-activity or hyper-activity patterns. Although PET data is naturally three-dimensional, we use one strategically chosen slice of our data ($Z=30$) as an exemplar as it cuts across regions of interest for AD diagnosis such as the hippocampus and temporal cortices \cite{da2017neuro}. 


\subsection{One-Sample Case} \label{subsec:one-sample}

\begin{figure}[h!] 
  \centering
  
  \quad\thinspace\textbf{Control Group}\par
  \begin{subfigure}[b]{0.05\linewidth}
    \includegraphics[width=\linewidth,height=4.4cm]{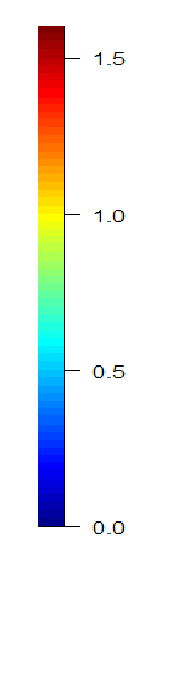}
  \end{subfigure}
  \begin{subfigure}[b]{0.3\linewidth}
    \includegraphics[width=\linewidth]{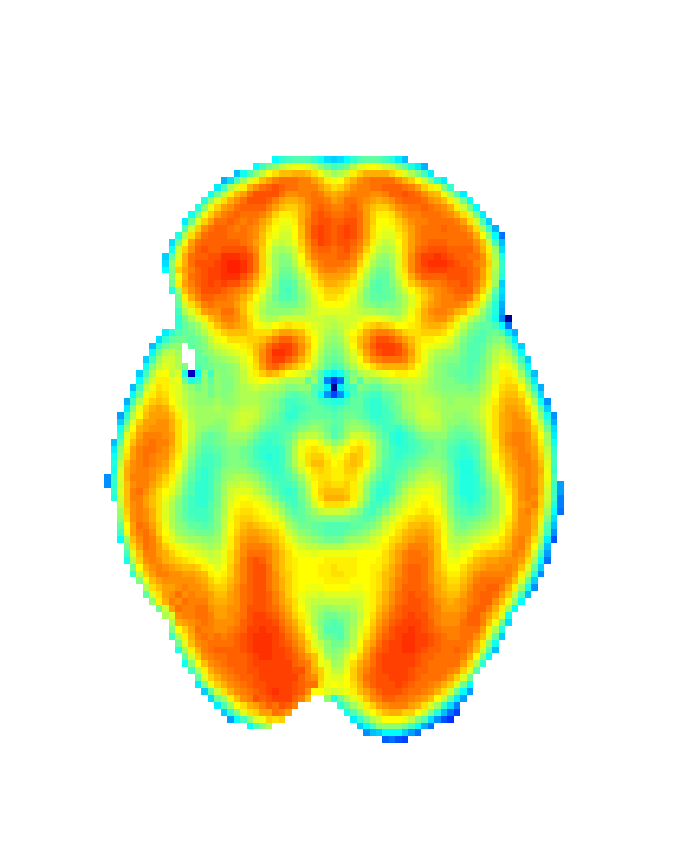}
    \caption{Lower SCCs}
  \end{subfigure}
  \begin{subfigure}[b]{0.3\linewidth}
    \includegraphics[width=\linewidth]{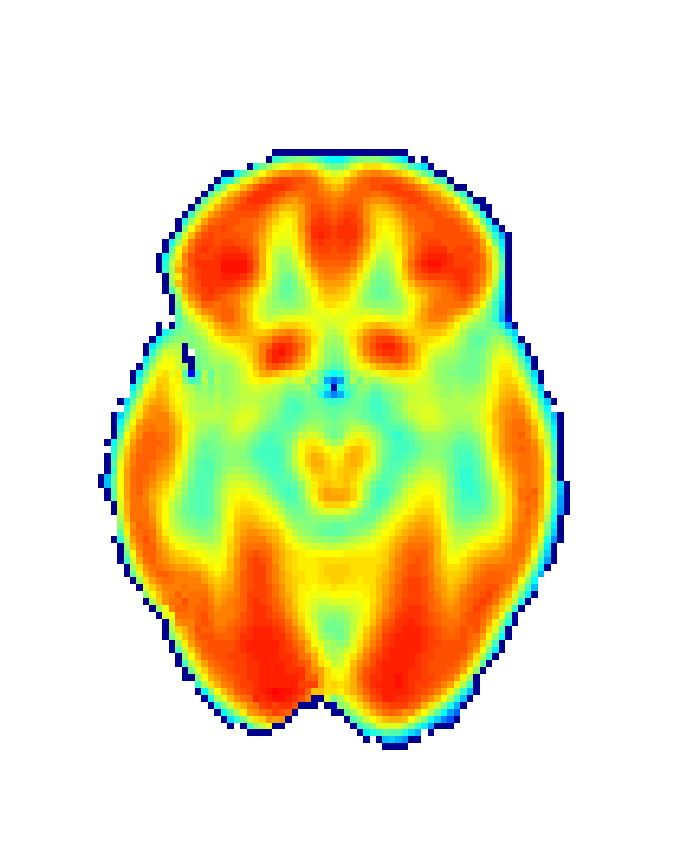}
    \caption{Mean Function}
  \end{subfigure}
  \begin{subfigure}[b]{0.3\linewidth}
    \includegraphics[width=\linewidth]{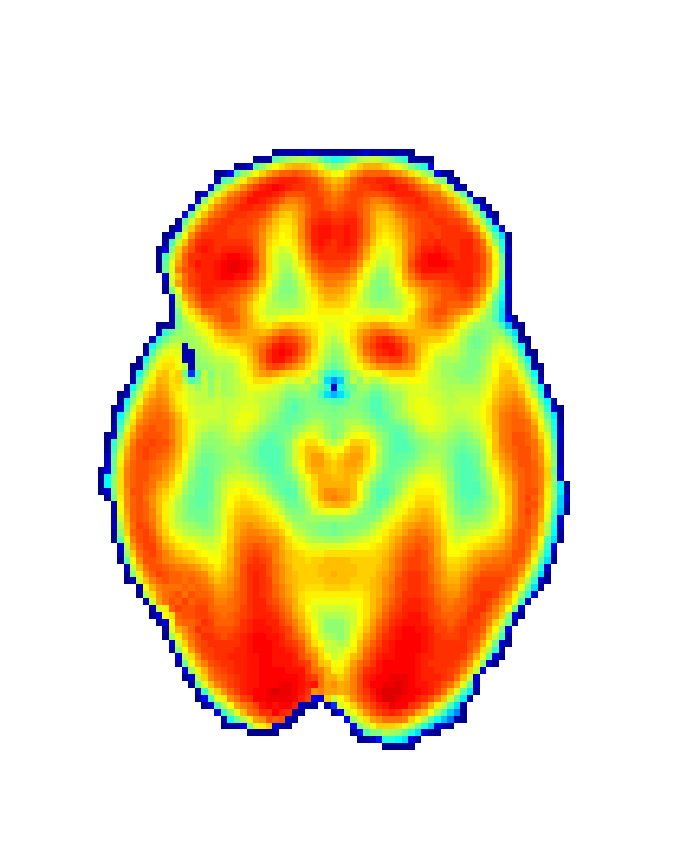}
    \caption{Upper SCCs}
  \end{subfigure}

  \vspace{0.5cm}
  \quad\thinspace\textbf{Alzheimer's Disease Group}\par    
  \begin{subfigure}[b]{0.05\linewidth}
    \includegraphics[width=\linewidth,height=4.4cm]{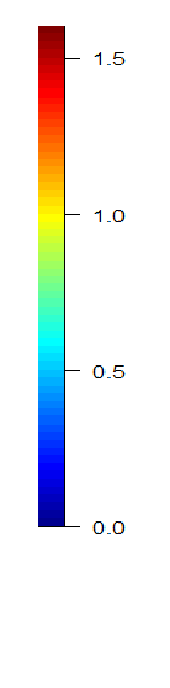}
  \end{subfigure}
  \begin{subfigure}[b]{0.31\linewidth}
    \includegraphics[width=\linewidth]{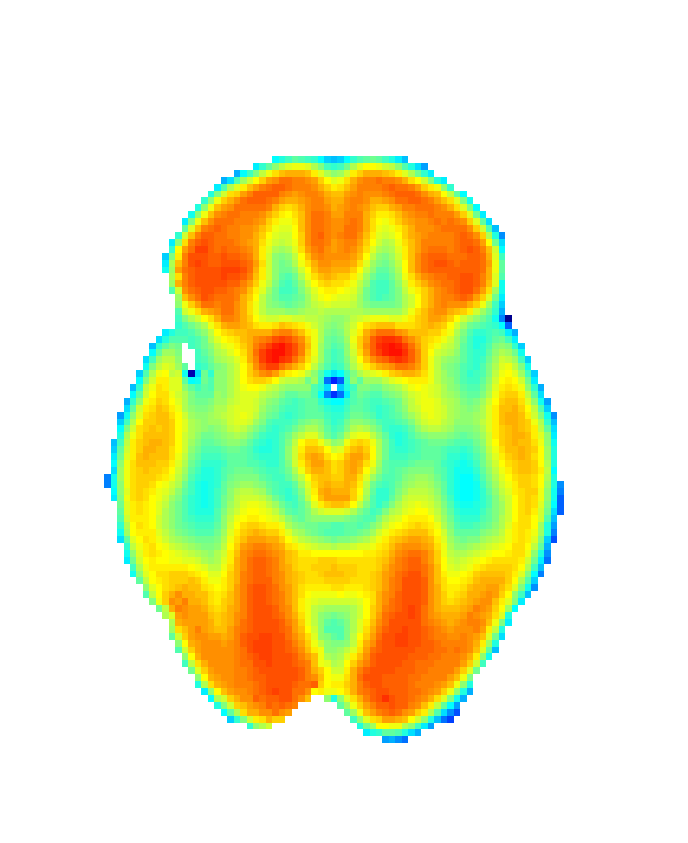}
    \caption{Lower SCCs}
  \end{subfigure}
  \begin{subfigure}[b]{0.3\linewidth}
    \includegraphics[width=\linewidth]{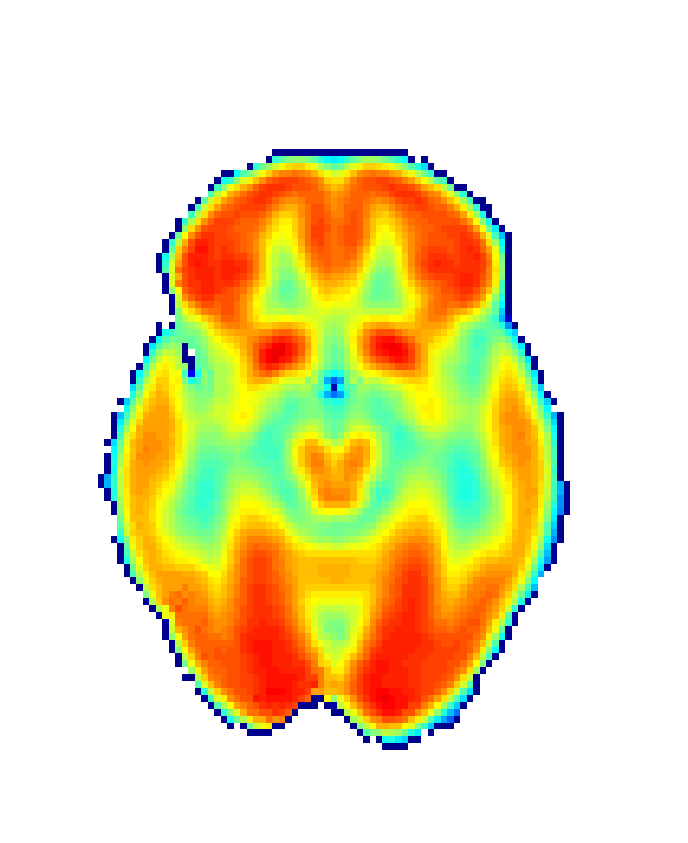}
    \caption{Mean Function}
  \end{subfigure}
  \begin{subfigure}[b]{0.3\linewidth}
    \includegraphics[width=\linewidth]{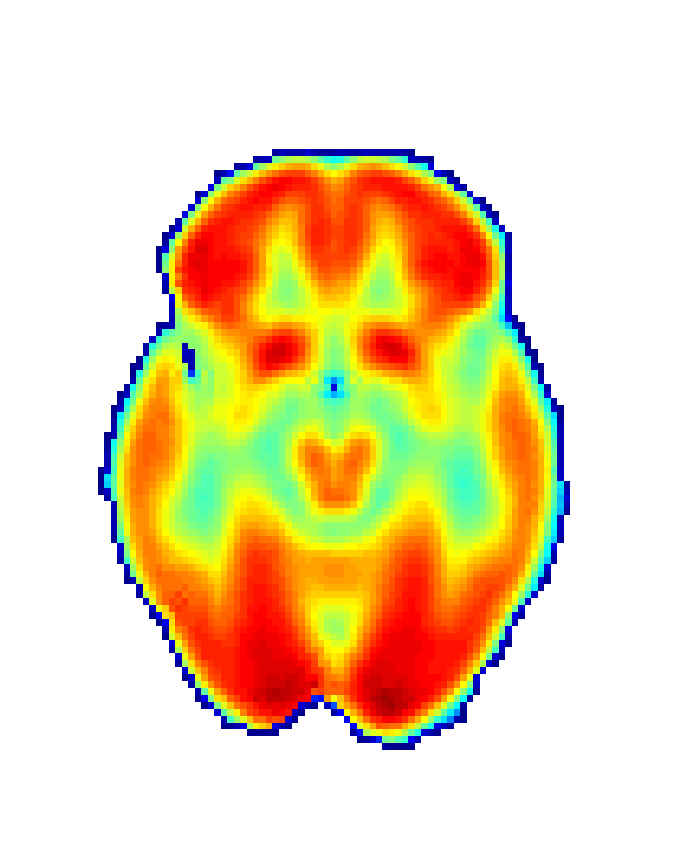}
    \caption{Upper SCCs}
  \end{subfigure}
 \vspace{0.5cm}
  \caption{Estimated mean function (center), lower SCCs (left), and upper SCCs (right) for CN group (n=75) and AD group (n=51). SCCs calculated for $\alpha=0.05$ using Delaunay triangulations (fineness degree  $N=8$).} \label{fig:MAIN_BASIC}
  \vspace{0.25cm}
\end{figure}

The first step to interpret our results is to consider the one-sample case. Figure \ref{fig:MAIN_BASIC} shows estimated mean function and SCCs ($\alpha=0.05$) for PET activity in CN and AD groups separately, providing a general overview of activity levels in a given two-dimensional slice. Although informative, these results have a limited utility as in real practice the aim is to examine differences between treatment groups, sexes, age blocks, or a combination of these. An experienced eye could already point out differences in activity levels from these images, however, the aim of this article is to provide the necessary tools to accurately detect brain regions which are suffering AD-derived atrophy; to achieve this end we now turn our attention to the two-sample case.

\vspace{0.5cm}


\subsection{Two-Sample Case} \label{subsec:two-sample}

For the two-sample case we estimate SCCs for the difference between CN and AD groups' mean functions, identify pixels whose PET values fall outside estimated SCCs (pinpointing abnormal activation patterns), and replicate the analysis for three different setups: entire group, filtering by sex, and filtering by age. We also display, as a visual assessment tool, results obtained using SPM under the same conditions.

Bear in mind that this methodology detects both hypo- and hyper-activation patterns. Finding hyper-activation patterns in AD brain regions compared to CN sounds counter intuitive at first. However, as we further discuss in Section \ref{sec:discussion}, there are different viable explanations for this finding (e.g., compensatory mechanisms, disturbances in normal brain deactivation processes). Nevertheless, we consider these findings are more likely attributable to a statistical artifact derived from the grand mean scaling and intensity normalization processes. For this reason we will only discuss results involving hypo-activity patterns in AD compared to CN groups.

\subsubsection{Entire Group} \label{subsubsec:entire-groups}

\begin{figure}[h!] 
  \centering
  \vspace{0cm}
    \textbf{Results using SPM}\par\medskip 
    \vspace{0cm}
  \begin{subfigure}[b]{0.32\linewidth}
    \includegraphics[width=\linewidth]{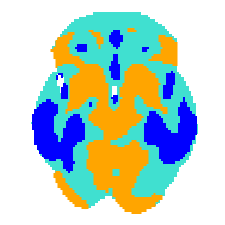}
    \caption{$\alpha=0.1$}
  \end{subfigure}
  \begin{subfigure}[b]{0.32\linewidth}
    \includegraphics[width=\linewidth]{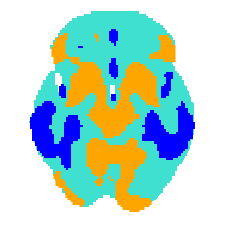}
    \caption{$\alpha=0.05$}
  \end{subfigure}
  \begin{subfigure}[b]{0.32\linewidth}
    \includegraphics[width=\linewidth]{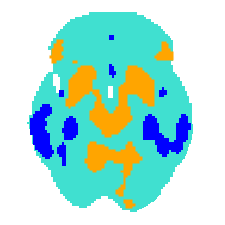}
    \caption{$\alpha=0.01$}
  \end{subfigure}
  \par\medskip
  \vspace{0.25cm}


 \textbf{Results using FDA}\par\medskip
  \begin{subfigure}[b]{0.32\linewidth}
    \includegraphics[width=\linewidth]{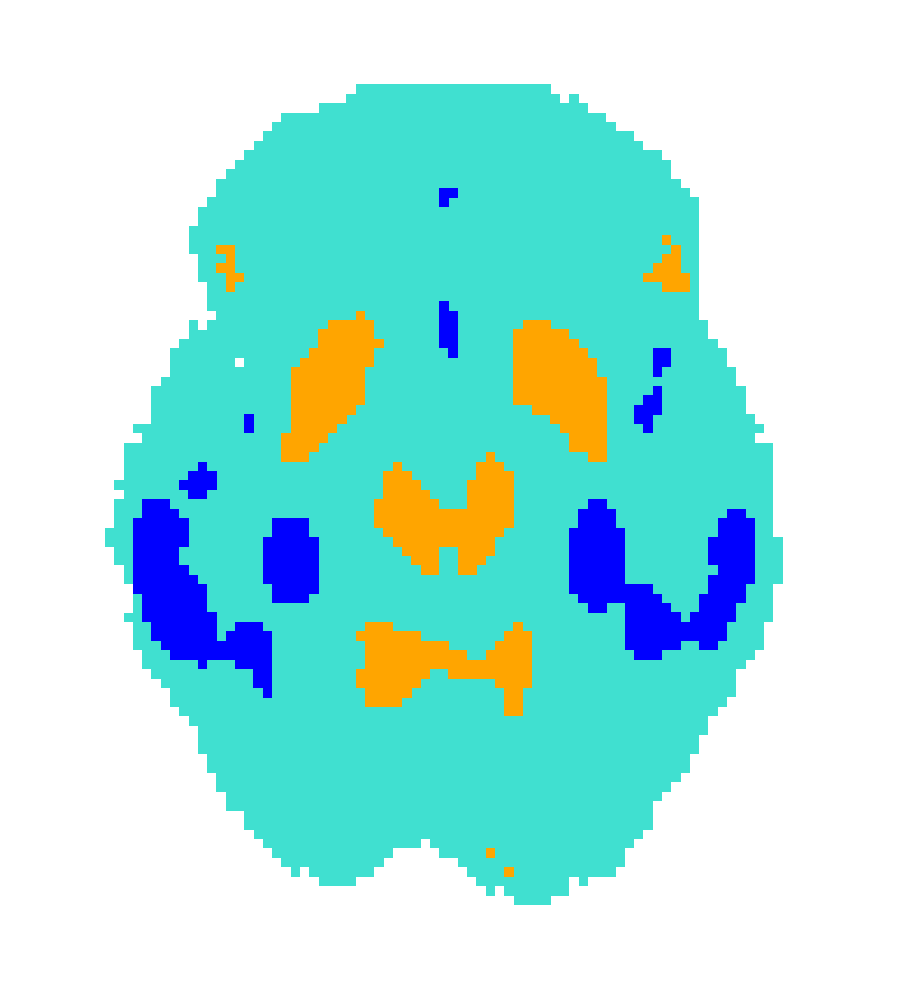}
    \caption{$\alpha=0.1$}
  \end{subfigure}
  \begin{subfigure}[b]{0.32\linewidth}
    \includegraphics[width=\linewidth]{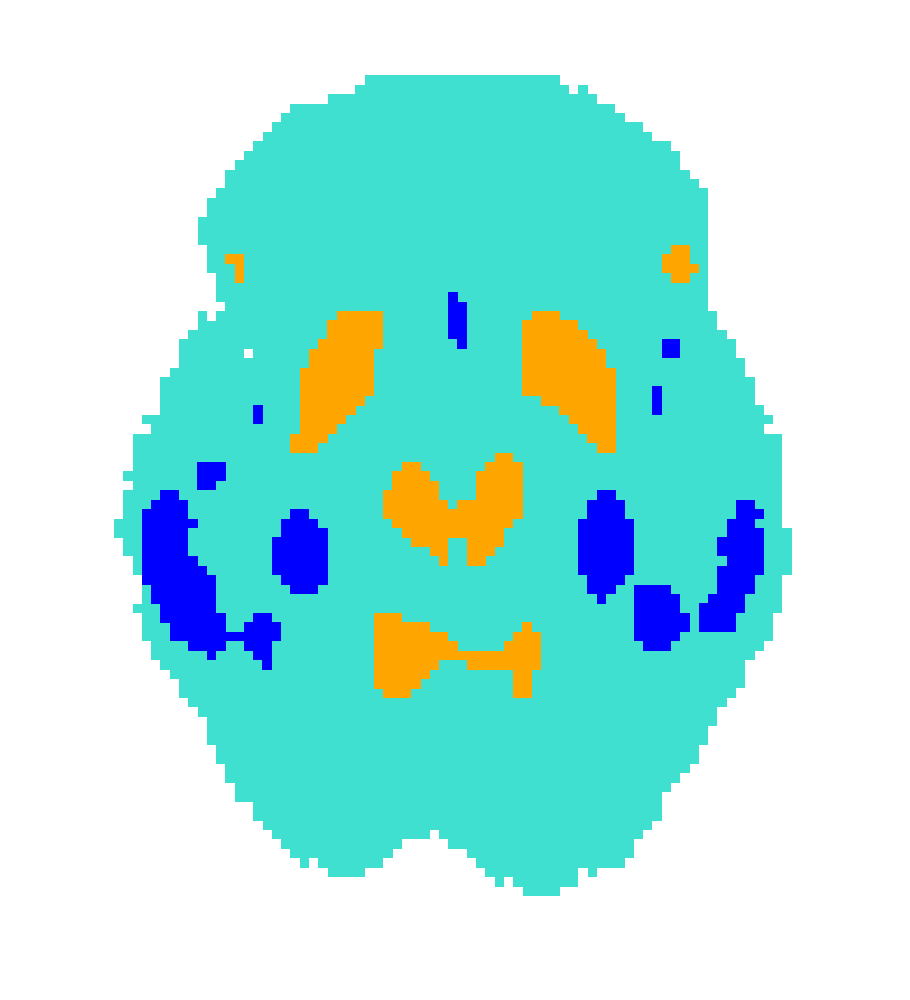}
    \caption{$\alpha=0.05$}
  \end{subfigure}
  \begin{subfigure}[b]{0.32\linewidth}
    \includegraphics[width=\linewidth]{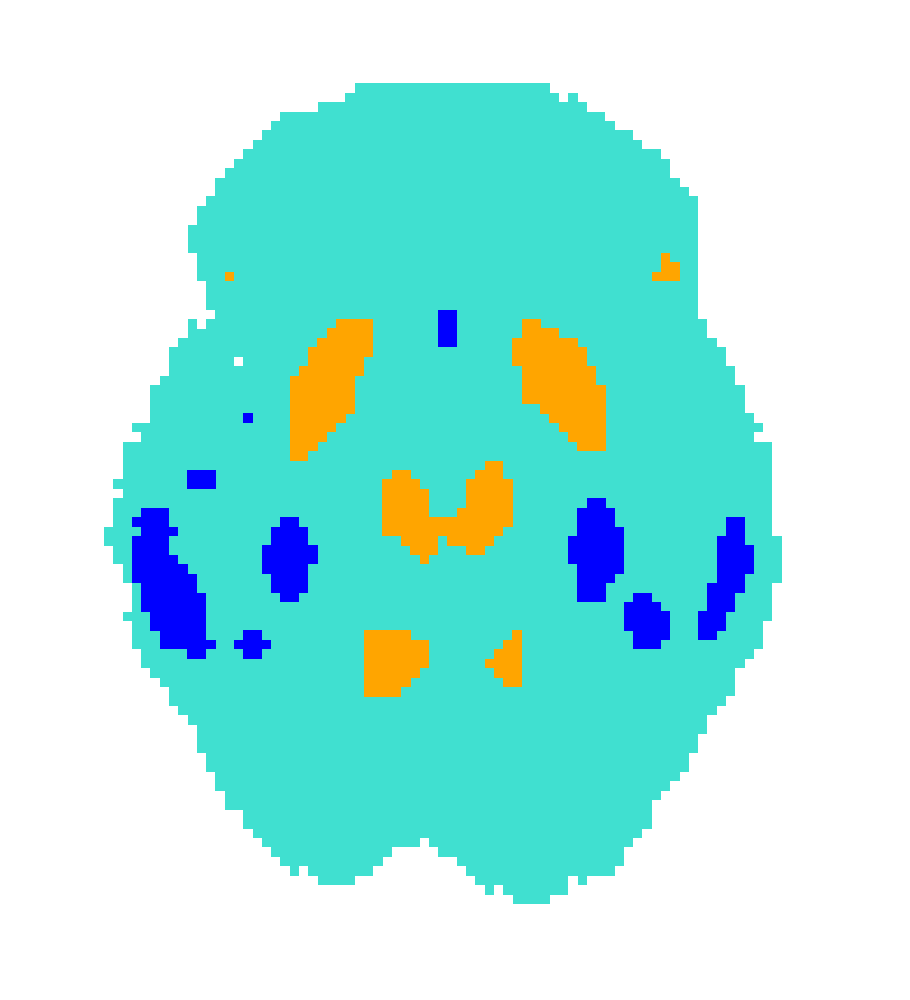}
    \caption{$\alpha=0.01$}
  \end{subfigure}
 \vspace{0.5cm}
  \caption{Comparison between classical SPM (uncorrected p-value) and FDA approaches for entire groups: AD (n=51) \textit{vs} CN (n=75) at different $\alpha$ levels. \textcolor{blue}{Blue} indicates hypo-activation in AD compared to CN while \textcolor{orange}{orange} indicates hyper-activation.} 
  \label{fig:TURQ_ALPHAS}
  \vspace{0.25cm}
\end{figure}

Figure \ref{fig:TURQ_ALPHAS} displays obtained results for a comparison between entire CN and AD groups. Upper row displays results using SPM, while lower row shows results obtained using our proposed FDA approach. Results are shown for decreasing $\alpha$ levels in both setups. This figure shows that, for any of the selected $\alpha$ levels, PET hypo-activity is found in temporal regions and hippocampus. This suggests that these regions are suffering a process of neural loss  attributable to AD pathology, findings which go in line with scientific literature on AD pathology \cite{da2017neuro}. Bear in mind that other regions which could be relevant for AD diagnosis - such as the parahippocampal gyrus and amygdala - are absent in the observed slice. 

Results using FDA and SPM show a strong resemblance in terms of regional distribution and point towards hypo-activation patterns in the same regions (i.e. temporal and hippocampal areas), although displaying different extensions. Besides, it is remarkable to observe that our FDA methodology appears to be more resilient to alterations in $\alpha$ levels than its counterpart, displaying results which are similar in distribution and extension across the three selected $\alpha$ levels. In contrast, SPM appears as a methodology whose results heavily depend on the selected $\alpha$ level.


\subsubsection{Filtering by Sex} \label{subsubsec:by-sex}

\begin{figure}[h!] 
  \centering
  \vspace{0.25cm}
  \textbf{Results using SPM}\par
  \vspace{0.35cm}
  \begin{subfigure}[b]{0.32\linewidth}
    \includegraphics[width=\linewidth]{figures/SPM_alpha/overlay005.png}
    \caption{Entire group}
  \end{subfigure}
  \begin{subfigure}[b]{0.32\linewidth}
    \includegraphics[width=\linewidth]{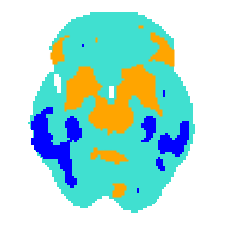}
    \caption{Female sub-group}
  \end{subfigure}
  \begin{subfigure}[b]{0.32\linewidth}
    \includegraphics[width=\linewidth]{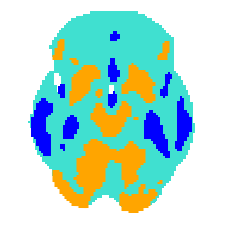}
    \caption{Male sub-group}
  \end{subfigure}
  \vspace{0.25cm}
  \vspace{0.25cm}
  \textbf{Results using FDA}\par
  \begin{subfigure}[b]{0.32\linewidth}
    \includegraphics[width=\linewidth]{figures/AD-CN/POINTS.png}
    \caption{Entire group}
  \end{subfigure}
  \begin{subfigure}[b]{0.32\linewidth}
    \includegraphics[width=\linewidth]{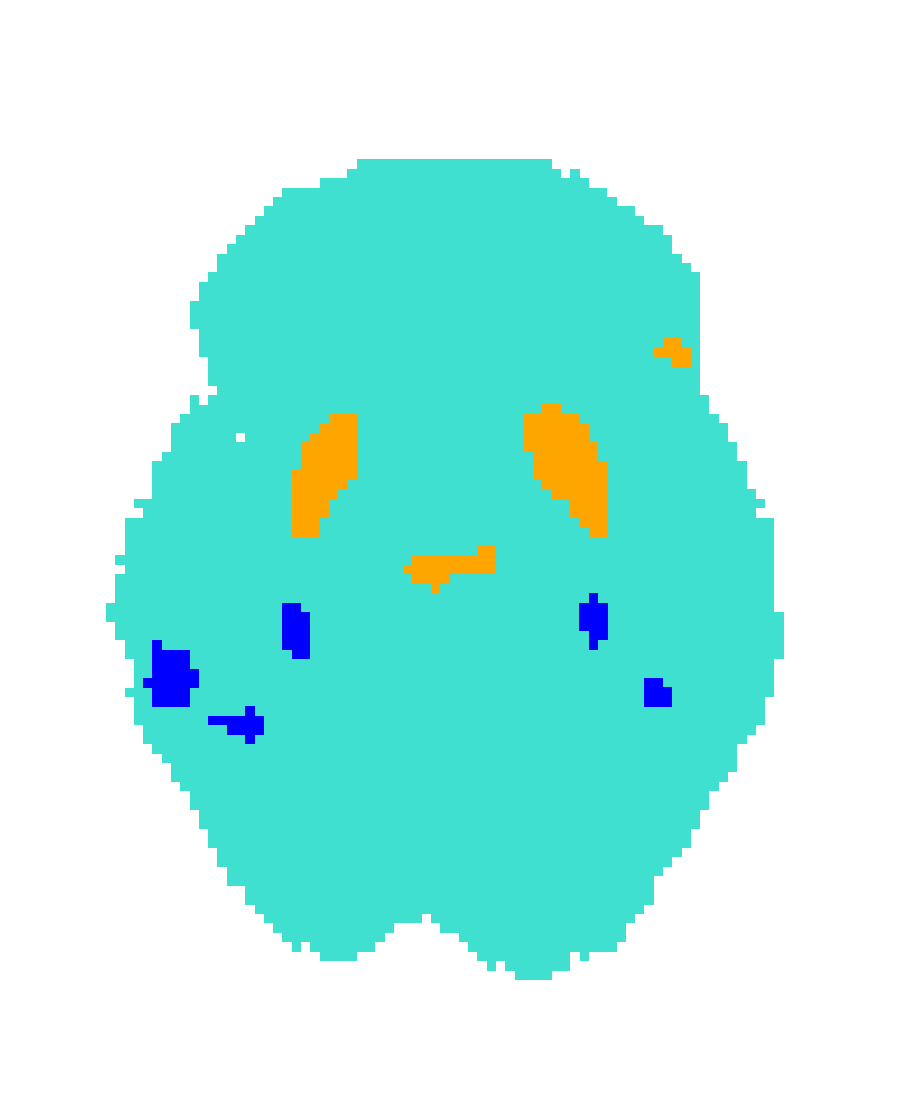}
    \caption{Female sub-group}
  \end{subfigure}
  \begin{subfigure}[b]{0.32\linewidth}
    \includegraphics[width=\linewidth]{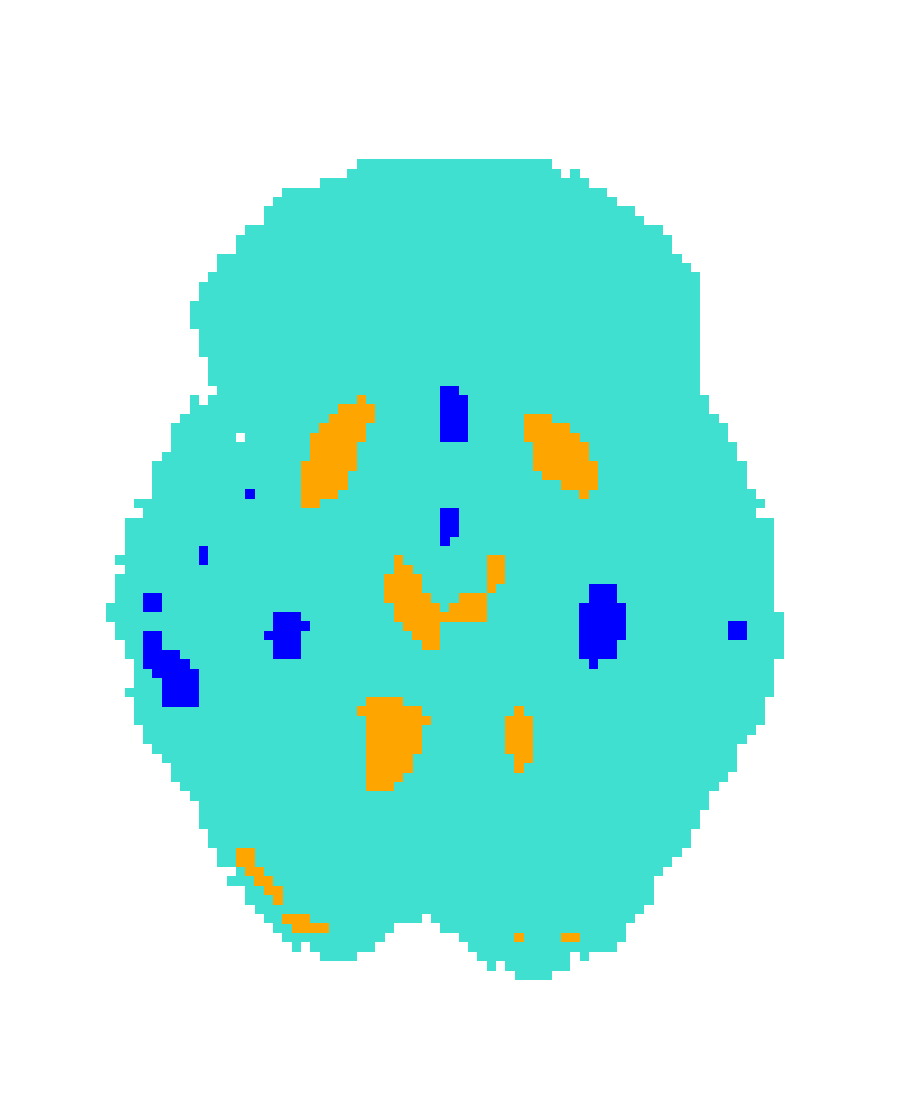}
    \caption{Male sub-group}
  \end{subfigure}
  \vspace{0.5cm}
  \caption{Comparison between SPM ($\alpha=0.05$; uncorrected p-value) and FDA approaches (SCCs for $\alpha=0.05$) for the analysis of groups filtered by sex. Female sub-group: CN (n=31) $\&$ AD (n=21); male sub-group: CN (n=43) $\&$ AD (n=30). \textcolor{blue}{Blue} indicates hypo-activation in AD compared to CN while \textcolor{orange}{orange} indicates hyper-activation.} 
  \label{fig:TURQ_SEX}
  \vspace{0.25cm}
\end{figure}

Figure \ref{fig:TURQ_SEX} shows results for the same comparison between AD and CN groups but now filtering by sex group. First, we display results for the entire group analysis as a visual reference. Following, we show results for SPM and FDA approaches filtering by female and male participants, aiming to identify differences in PET activity between AD and CN groups filtered by sex. In this figure we can see that obtained results for SPM are much more diffuse than the ones obtained using an FDA approach. SPM results display large regions of hypo-activity both for female and male sub-groups which can be identified as parts of the hippocampus and temporal cortices. This noisy response is easily identified, for example, in Figure 6.b where hypo-activity is found in an area that goes from left temporal to left occipital lobes; this result does not go in line with the available literature on AD pathology and is likely attributable to a poor analysis. 

On the other hand, results obtained using our FDA approach are more cautious and, although the distribution of regions identified as hypo-active remains similar to the ones identified with SPM, the extension of these areas is much more reduced compared to its counterpart, with hypo-activity nuclei clearly located in temporal and hippocampal regions both for female and male sub-groups and identifying a much less numerous amount of sparse nuclei. 

Overall, both methods present results which go in line with the available literature (i.e. there are no apparent differences in regional distribution of neural loss between female and male AD patients) but our FDA approach appears to be more cautious in the displayed results following reductions in sample size. This is easily observed in the current figure where, although regional distributions of PET hypo-activity remain similar across methodologies, the extension of these areas is much more constrained in our proposed FDA setup.



\subsubsection{Filtering by Age} \label{subsubsec:by-age}



\begin{figure}[h!] 
  \centering
  \vspace{0.25cm}
  \textbf{Results using SPM}\par
  \vspace{0.35cm}
  \begin{subfigure}[b]{0.32\linewidth}
    \includegraphics[width=\linewidth]{figures/SPM_alpha/overlay005.png}
    \caption{Entire Group}
  \end{subfigure}
  \begin{subfigure}[b]{0.32\linewidth}
    \includegraphics[width=\linewidth]{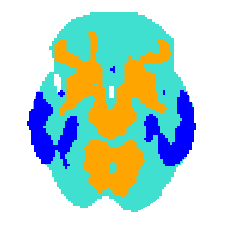}
    \caption{$\leq$ 75 y.o. sub-group}
  \end{subfigure}
  \begin{subfigure}[b]{0.32\linewidth}
    \includegraphics[width=\linewidth]{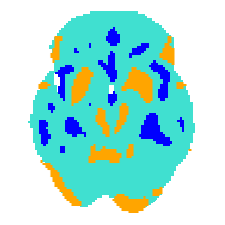}
    \caption{$>$75 y.o. sub-group}
  \end{subfigure}
  \par\medskip
  \vspace{0.25cm}
  \textbf{Results using FDA}\par
  \begin{subfigure}[b]{0.32\linewidth}
    \includegraphics[width=\linewidth]{figures/AD-CN/POINTS.png}
    \caption{Entire group}
  \end{subfigure}
  \begin{subfigure}[b]{0.32\linewidth}
    \includegraphics[width=\linewidth]{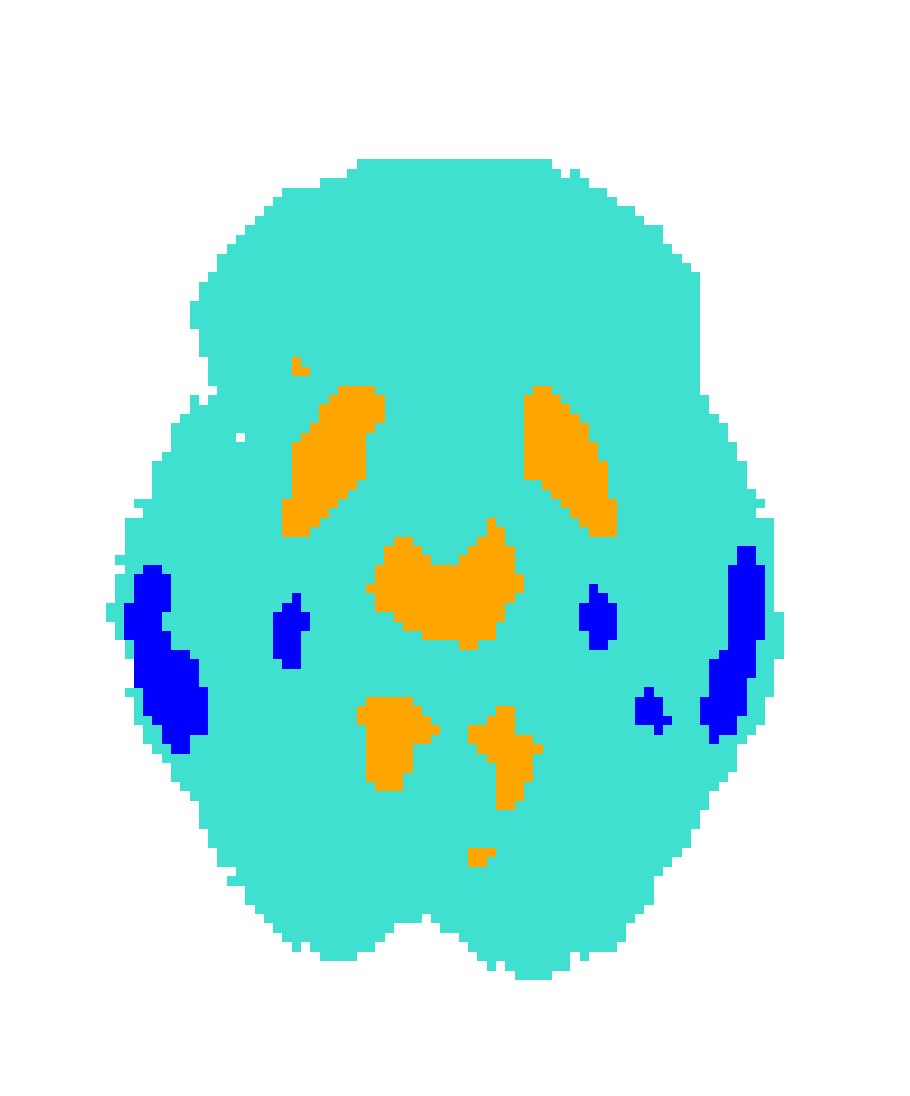}
    \caption{$\leq$ 75 y.o. sub-group}
  \end{subfigure}
  \begin{subfigure}[b]{0.32\linewidth}
    \includegraphics[width=\linewidth]{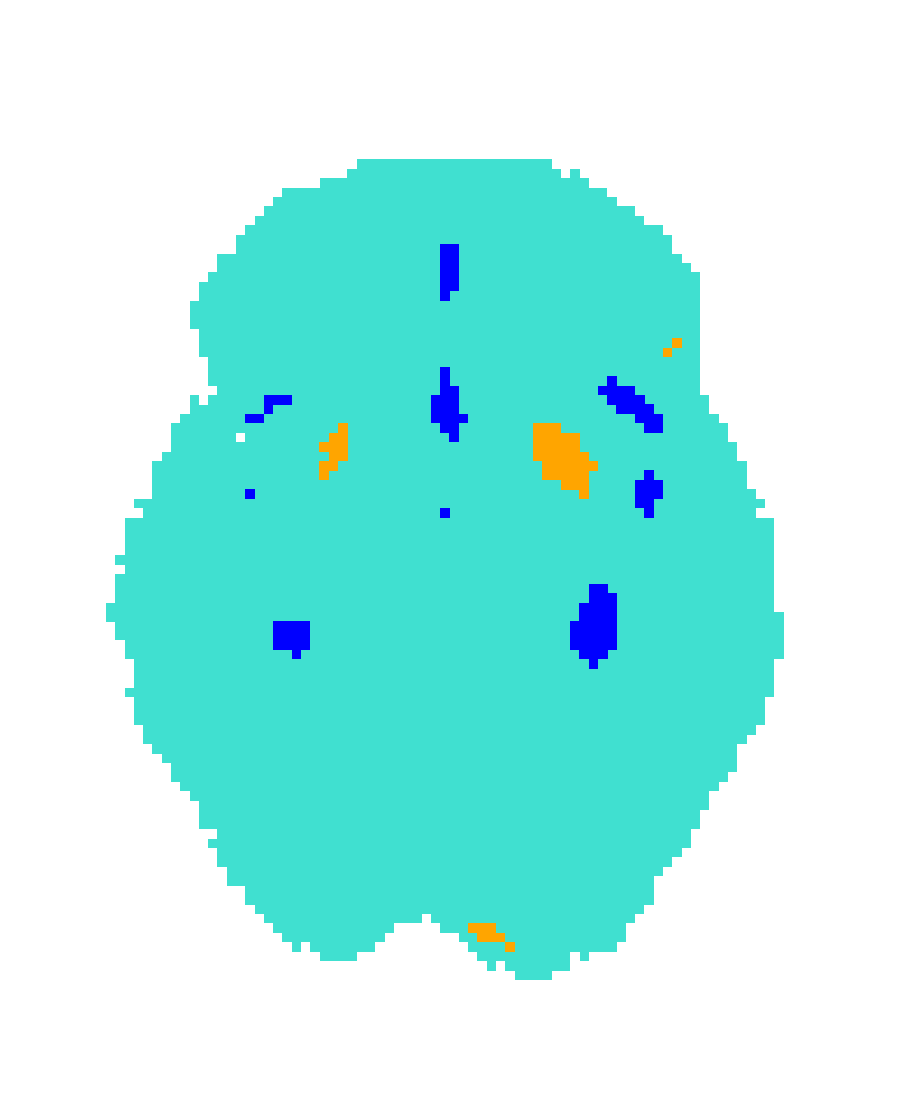}
    \caption{$>$75 y.o. sub-group}
  \end{subfigure}
  \vspace{0.5cm}
  \caption{Comparison between SPM ($\alpha=0.05$; uncorrected p-value) and FDA approaches (SCCss for $\alpha=0.05$) for the analysis of groups filtered  by age. $\leq$ 75 y.o. sub-group: CN (n=31) $\&$ AD (n=21). $>$75 y.o. sub-group: CN (n=43) $\&$ AD (n=30). \textcolor{blue}{Blue} indicates hypo-activation in AD compared to CN while \textcolor{orange}{orange} indicates hyper-activation.} 
  \label{fig:TURQ_AGE}
  \vspace{0.25cm}
\end{figure}


Figure \ref{fig:TURQ_AGE} shows results for the same comparison between AD and CN groups filtering by age block. First, we display results for the entire group analysis as a visual reference. Following, we filter by age, dividing our populations in a $\leq$ 75 y.o. sub-group and a $>$75 y.o. sub-group. We perform this analysis in duplicate, both for SPM and for FDA approaches under the same setup. This figure shows patterns of hypo-activation which are similar between SPM and FDA methodologies, pointing towards reductions in PET activity in hippocampal and temporal regions, a pattern which is more easily identified in the $\leq$ 75 y.o. sub-group. SPM results are more diffuse and highlight large regions and numerous nuclei showing hypo-activation. On the other hand, our FDA approach is again more cautious, with reduced extensions and scarcer nuclei identiﬁed for the $\leq$ 75 y.o. sub-group. The same happens with the $>$75 y.o. sub-group but, in this case, our FDA approach identiﬁes a reduced number of hypo-activity nuclei where SPM identiﬁes numerous and scattered nuclei which are difficult to interpret. 

Further, both methods’ results - although similar, especially for the $\leq$ 75 y.o. sub-group - are at ﬁrst counter intuitive as it appears that the expected PET activity reductions in temporal regions and hippocampus are only present in the $\leq$ 75 y.o. sub-group, but disappear for the $>$75 y.o. sub-group. However, it can be argued that differences in brain activity between CN and AD groups are more intense in early stages of life, as opposed to older populations where normal ageing features overlap and create a confusion effect with the properly pathological features of AD. In addition, obtained results might also suffer from distortions due to reductions in sample size.


\section{Discussion} \label{sec:discussion}

This article focuses on the growing medical problem of AD in our societies, considered as one of the greatest challenges for neuroscience in this century and with estimates predicting more than a hundred million patients worldwide by 2050 \cite{world2015world}. Considering that AD has no known cure and the limited success of therapeutic interventions, the most promising field of study is early AD diagnosis. This is supported by research suggesting that AD may begin between 20 to 30 years prior to first observable symptoms \cite{Goedert2006}. Thus, early identification and treatment appear as a critical step, especially bearing in mind that, as an age-related condition, small delays on disease onset can significantly reduce AD prevalence \cite{Briggs2016}. 

In this article we intend to improve AD diagnostic tools by comparing classical diagnostic methods with a new proposal whose theoretical basis were recently settled by Wang et al. \cite{Wang2019}. This new approach avoids the multiple testing cumulative errors problem, performs well when dealing with complex data structures, and overall represents a step forward in the way we perform statistical analysis of data in the medical imaging field. Our aim for this article is to prove by means of visual assessment that results obtained using this novel approach are at least equally valid for AD diagnosis as the ones obtained using traditional approaches. Further, we also intend to prove that this methodology might be more resilient than SPM in analysis that implicate complex structures and big amounts of data.   

It is worth noting that our analysis detects both hypo and hyper-activations in the difference between groups. As a result, although counter intuitive, our result figures also display regions with increases in PET activity in AD compared to CN (e.g., putamen, vermis, cerebellum, some cortical regions). These results are not rare in the neuroimaging literature; previous publications have explained them suggesting that AD may produce compensatory changes in neural activity \cite{schwindt2009functional} or that hyperactivity is actually a sign of disturbance in normal deactivation processes \cite{bejanin2012higher}. However, even though there is evidence suggesting the existence of compensatory mechanisms in AD and disturbances in normal deactivation patterns, we consider that this effect in our results is more likely attributable to a statistical artifact caused by the process of grand mean scaling and intensity normalization. For this reason we only evaluate results involving hypo-activity patterns in AD compared to CN groups.

We carried out this process of regional identification for the difference between AD and CN estimated mean functions, first for entire groups and then filtering by sex and age blocks. For the entire group analysis, Figure \ref{fig:TURQ_ALPHAS} shows hypo-activity patterns in temporal and hippocampal regions for any of the selected $\alpha$ levels both for traditional SPM and novel FDA approaches. Albeit results were similar, it appears to us that the proposed FDA methodology is more cautious and more resilient to changes in $\alpha$ levels than its SPM counterpart. Besides, Figure \ref{fig:TURQ_SEX} and Figure \ref{fig:TURQ_AGE} demonstrate that the proposed FDA approach is also more resilient to reductions in sample size than classic SPM approaches, displaying reduced nuclei of hypo-activation both in terms of quantity and size. This is especially easy to visualize in Figure \ref{fig:TURQ_AGE}.c and Figure \ref{fig:TURQ_AGE}.f where SPM identifies a big number of sparse and small nuclei showing patterns of hypo-activity while, on the other hand, the proposed FDA approach identifies a more reduced number of brain areas as relevant. Overall, results for both methodologies differ in terms of the extension of identified areas, but not in terms of distribution of these areas. This points out towards the validity of our proposed FDA approach, which seems to identify the same areas suffering from AD-derived neural loss than SPM does.

In summary, our goal for this article was to implement Wang et al.'s \cite{Wang2019} methodology from a theoretical framework into a practical case in an attempt to demonstrate that this novel methodology is valid for AD diagnosis and potentially other medical imaging analysis. After visually assessing results displayed in Figure \ref{fig:TURQ_ALPHAS}, Figure \ref{fig:TURQ_SEX}, and Figure \ref{fig:TURQ_AGE} we consider our initial goal for this article as a success and acknowledge this statistical approach's potential in the field of medical imaging. In addition, this FDA approach circumvents the multiple comparison problem and appears to display a better resistance to changes in $\alpha$ levels and reductions in sample size than classic SPM. However, there is still much work to do in order to refine this methodology and to apply it in routine clinical practice.


\section{Limitations \& Further Research} \label{sec:limitations}

Although promising, the results obtained using this FDA novel technique suffer from intrinsic limitations derived from the proposed setup and available data. We used this methodology with the aim of detecting brain atrophy basing our results on 18FDG-PET data, however, inappropriate use of 18FDG-PET anatomic standardization processes can cause a series of statistical artifacts and systematic biases which have to be taken in account to avoid a serious problem of false positives \cite{ishii2001statistical,Lopez-Gonzalez2020}. 

Besides, this technique has been applied to a limited number of participants (126 participants: 51 AD \& 75 CN). Future research should increase the number of participants and also increase the number of conditions considered by including transitional stages between healthy and AD populations such as patients suffering from MCI. Additionally, as we stated previously in this article, neuroimaging data is naturally three-dimensional. In this article we work with a selected slice of two-dimensional data which is especially relevant as several brain regions involved in AD diagnosis cut through it. However, future research should search for a way of extending this methodology to the three-dimensional case without losing estimator's consistency. This would be a great step forward for neuroscience and medical imaging in general, albeit not without lots of effort both in theoretical and computational terms. 

It would also be of great interest to implement this technique into a single-participant setup, as in clinical practice the most useful approach for diagnosis consists on comparing one single patient with data from a large group of CN patients in order to visualize which brain areas show different patterns of activation and thus decide whether the patient suffers from AD or not. However, the application of this methodology to a single-case setup has been so far unsuccessful. In addition, the herein proposed methodology would provide a new tool for clinicians to make grounded diagnosis. However, future research and computational effort should strive for the implementation of an automatic classification system based on SCCs for neuroimaging data. Such a system would not only provide the clinical specialist with better images to make a decision but also with an automatic proposed classification (either as healthy or pathological) based on large datasets of previous patients. 

Furthermore, although the utility of this technique is confirmed visually by the great resemblance between considered approaches, future research should strive for fully establishing SCCs as an alternative to traditional methodologies by mathematically demonstrating its performance compared to SPM in different setups. Finally, as we mentioned before in this article, the proposed methodology is not limited to AD research but rather the opposite, as it is potentially extensible to any other neurological diseases and, overall, to any other project involving medical imaging in which neural activity and/or volume calculations are of clinical interest. Future research should aim for the implementation of this technique to other research projects in order to fully elucidate the potential of this novel approach.


\clearpage

\section*{Acknowledgments}

The authors are grateful for the collaboration and input provided by several members of the \href{https://grid-usc.com/equipo}{GRID-BDS} team. Neuroimaging data was extracted from the ADNI database which is curated by a \href{http://adni.loni.usc.edu/wp-content/uploads/how_to_apply/ADNI_Acknowledgement_List.pdf}{series of researchers} to whom, even though they did not participate in the elaboration of this research paper, we are very grateful. Juan A. Arias also wants to acknowledge the passive support of his radio partner \href{https://www.instagram.com/uberflut/?hl=es}{@\textit{uberflut}}. May the red light bulb shine forever in their studio. 

%
%
%

\bibliographystyle{splncs04}
\bibliography{biblio}
\end{document}